Potential Benefits of Employing Large Language Models in Research in Moral Education

and Development

Hyemin Han[1]

[1] Educational Psychology Program, University of Alabama

Author Note

We have no known conflict of interest to disclose.

Correspondence concerning this article should be addressed to Hyemin Han,

University of Alabama, Box 872031, Tuscaloosa, AL 35487, United States.

Email: hyemin.han@ua.edu

Potential Benefits of Employing Large Language Models in Research in Moral Education
and Development

**Abstract**

Recently, computer scientists have developed large language models (LLMs) by
training prediction models with large-scale language corpora and human reinforcements.
The LLMs have become one promising way to implement artificial intelligence with
accuracy in various fields. Interestingly, recent LLMs possess emergent functional features
that emulate sophisticated human cognition, especially in-context learning and the chain of
thought, which were unavailable in previous prediction models. In this paper, I will
examine how LLMs might contribute to moral education and development research. To
achieve this goal, I will review the most recently published conference papers and ArXiv
preprints to overview the novel functional features implemented in LLMs. I also intend to
conduct brief experiments with ChatGPT to investigate how LLMs behave while addressing
ethical dilemmas and external feedback. The results suggest that LLMs might be capable of
solving dilemmas based on reasoning and revising their reasoning process with external
input. Furthermore, a preliminary experimental result from the moral exemplar test may
demonstrate that exemplary stories can elicit moral elevation in LLMs as do they among
human participants. I will discuss the potential implications of LLMs on research on moral
education and development with the results.

exemplar*, *Simulation*

**Introduction**

One of the most impactful recent developments in computer science is large language models (LLMs) (Grossmann et al., 2023), which implement advanced artificial intelligence. Computer scientists developed LLMs to predict the most probable solution to a given inquiry by training prediction models with large amounts of language input (Zhao et al., 2023). LLMs utilize large-size corpora from various sources to train their prediction models. Now, people working in fields other than computer science are using such LLMs for various purposes. For instance, ChatGPT based on one of the most widely used LLMs, i.e., GPT, has significantly influenced every aspect of human lives due to its user-friendliness and versatility (Mogavi et al., 2023). People employ ChatGPT to achieve diverse goals, such as drafting and reviewing paragraphs, preprocessing data, and generating source codes for computer programing. They found that ChatGPT and general LLMs can produce high-quality products following directions provided in human languages (Dwivedi et al., 2023).

Because computer science evolves rapidly, I will briefly overview recent conference papers and ArXiv preprints addressing LLMs to examine how they work and their distinctive functional features. Computer scientists invented LLMs with large-scale language datasets to make them capable of understanding human language input and producing plausible responses (Zhao et al., 2023). They develop LLMs by training prediction models with large-scale corpora, which include demonstrations and exemplars. Scientists link the exemplar inputs (X) with labels indicating whether or to what extent a specific X is desirable (Y) while training LLMs. In the long run, LLMs tend to produce a predicted output (ŷ) in a given condition (x) that is most likely and deemed most desirable based on the trained set. Because LLMs include functionalities to interpret natural language input during the process, they can understand x and generate $\hat{y}$ in human language (Arcas & Agüera, 2022). When they generate $\hat{y}$ with low quality, it is possible to provide feedback to LLM so that they correct their prediction models through human reinforcement learning (Ouyang et al., 2022; Srivastava et al., 2023). For instance, when LLMs generated biased statements against specific groups, providing human feedback reduced the bias significantly (Ganguli et al., 2023).

Such training and learning processes are similar to what occurs among humans. The abovementioned processes induced by demonstrations and exemplars resemble Bayesian learning at the behavioral and neural levels (Friston, 2003). Each individual has prior beliefs existing before experiences. External inputs, such as observations, examples, and instructions, update the existing priors into posteriors (Mathys, 2011). Although the Theorem does not predict the posterior update completely, the mechanism of Bayesian learning well approximates human learning processes (McDiarmid et al., 2021). Recent works proposed that the Bayesian learning mechanism can also apply to moral psychology and development (Cohen et al., 2022; Han, 2023a; Railton, 2017).

Interestingly, recent developments in LLMs demonstrate significant emerging features simulating human psychological processes, which were unavailable in the previous simulation models (Grossmann et al., 2023). I would like to review two major emerging features: in-context learning and reasoning and the chain of thought and reasoning that provide greater degrees of freedom and flexibility in cognition in simulation. First, the most up-to-date LLMs can conduct in-context learning (Dong et al., 2023). Unlike simple machines designed to provide specified responses to specified inquiries, LLMs can learn from a set of contextually relevant exemplars and demonstrations (Dong et al., 2023;

Zhao et al., 2023). Then, they can solve problems within a similar context even if they are not directly identical to the examples used for training. The in-context learning is a feature that emerged from the large-scale and complex network constituting LLMs, which was not available in the past when available computational resources did not allow the implementation of large-size prediction models (Dong et al., 2023). At this point, in the domains of learning and problem-solving, LLMs can behave more similarly to human beings than the previous simulation models because they can exercise contextual training and prediction with enhanced flexibility (Moor et al., 2023; Wu et al., 2023).

Second, recent LLMs show emerging capabilities of the chain of thought and reasoning (Wei et al., 2023). Previously, computers could only provide direct answers to problems without elaboration on the process of problem-solving (Zhao et al., 2023). For instance, once we ask, "Corgi has two retriever puppets. She purchased two more boxes with two puppets in each. How many retriever puppets do Corgi have?" they could say "8" without explaining any rationale. Unlike the previous computational models, LLMs can explain the process of thought and reasoning and can even learn from the explanation of such a process provided by humans. In the case of the abovementioned inquiry, LLMs can explain their rationale, such as "Corgi started with two puppets. Three boxes of two puppets each are six puppets. 2 + 6 = 8." Providing the chain of reasoning significantly improved LLMs' performance in math problem-solving (Wei et al., 2023). Likewise, in the case of the bias correction study, when researchers provided additional feedback accompanying the chain of reasoning, LLMs showed significantly more decreased bias than when they gave the models simple factual corrections (Ganguli et al., 2023). Like humans, LLMs can improve their prediction models more effectively by learning from the examples of thinking and reasoning processes (Huang & Chang, 2023; Zhao et al., 2023). In addition to the capability of in-context learning, the chain of thought capability suggests that recent LLMs can exercise rudimentary forms of reasoning with contextual flexibility (Huang & Chang, 2023).

Such advanced emerging features of LLMs suggest LLMs might be able to simulate rudimentary human moral functioning. Related to human morality, several previous studies examined whether LLMs are capable of morality-related cognitive faculties, such as the theory of mind (ToM) and reasoning (May, 2018; Young et al., 2007), thanks to the newly emerging features. Although the evidence is equivocal (Shapira et al., 2023), some demonstrated that most up-to-date LLMs showed rudimentary forms of ToM equivalent to what seven-year-olds may perform (Kosinski, 2023). One study reported that ChatGPT could generate philosophical statements mimicking those by Daniel Dennett, an experimental philosopher, that non-expert participants could not accurately distinguish from his original ones (Schwitzgebel et al., 2023). Furthermore, one previous study examined whether LLMs can correct their bias against marginalized groups (Ganguli et al., 2023). In the study, when researchers provided the models with feedback for correction, LLMs reported a significant decrease in discrimination against minority groups. Of course, I admit that these are insufficient to support arguments that LLMs are fully equipped with the cognitive faculties that humans possess, so whether we should consider them moral agents and whether they can perform moral functioning as human beings are still controversial (Chalmers, 2023). Hence, I intend to focus on the potential practical values of LLMs while considering their current technological limitations.

**Current Paper**

In this paper, I propose that LLMs can eventually assist our research on moral education and development, particularly those involving empirical and practical investigations notwithstanding current limitations. I will examine how LLMs can contribute to research in moral education. Philosophers are primarily interested in addressing the potential ethical impacts of LLMs or how enhanced artificial intelligence based on LLMs will change philosophical inquiries about cognition and psychology (e.g., Hosseini et al., 2023). In educational research, researchers started considering the educational implications of LLMs, particularly those related to teaching and learning, such as how they will change ways to teach and learn writing, search for information, etc. (e.g., Kasneci et al., 2023; Milano et al., 2023). Unlike these previous works, in this paper, I will explicitly focus on how the development of LLMs will influence empirical and practical research in moral education with concrete points. Hence, the abovementioned issues related to the ethical and educational implications of LLMs are out of the scope of my paper.

In the first section, I will briefly test LLMs' capability to conduct moral reasoning with ethical dilemmas. Then, I will examine whether LLMs' in-context learning and reasoning and the chain of thought and reasoning features can facilitate revisions of moral reasoning with contextual information and feedback, which are closely associated with moral educational activities. Furthermore, I will also examine whether the stories of moral exemplars can elicit emotional and motivational impacts in LLMs. Because moral reasoning alone cannot accurately explain the mechanism of moral motivation and behavior (Kristjánsson, 2010), I planned to employ the moral exemplar intervention, which primarily targets moral emotion and motivation; moral educators tend to utilize the intervention to generate affective and motivational impacts rather intuitively (Kristjánsson, 2017; Sanderse, 2012). Given demonstrations constitute the basis for the learning mechanisms of LLMs (Zhao et al., 2023), I expected that this moral exemplar intervention might also produce similar outcomes in LLMs as among human participants. Based on the concrete experimental outcomes, I intend to discuss the implications of LLMs on research in moral education and development. Finally, I will overview the limitations and future directions with concluding remarks.

## The Behavioral Defining Issue Test

I examined how LLMs address moral problems to acquire insights about whether they can exercise some aspects of moral functioning, especially moral reasoning, like humans, based on the abovementioned functionalities, i.e., in-context learning, and the chain of thought and reasoning. Thus, I briefly tested how ChatGPT (May 24 version) solves ethical dilemmas in the behavioral Issues Test (bDIT). The bDIT is a simplified version of the DIT that assesses individuals' development of moral reasoning in terms of postconventional reasoning (Choi et al., 2019; Han, Dawson, et al., 2020). I employed the bDIT instead of the original DIT due to its simple structure, which can be feasibly implemented in the ChatGPT environment. Because I used the free ChatGPT (https://chat.openai.com/), I assumed that only the general corpora, which were not specific about moral philosophy and psychology, were used to train the GPT model (Guo et al., 2023).

First, I entered the dilemmas and items asking for moral philosophical rationale supporting behavioral decisions quoted from the bDIT. The bDIT presents three dilemma stories, i.e., Heinz and the drug, Newspaper, and Escaped Prisoner. For each dilemma, I asked ChatGPT to evaluate whether a provided behavioral solution (e.g., "Should Heinz steal the drug?" in the case of Heinz and the drug) was morally appropriate. Then, I asked eight questions about the most important rationale supporting the decision per dilemma. For each question, I presented three options corresponding to three reasoning schemas, i.e., personal interest, maintaining norms, and postconventional schemas (Rest et al., 1999). Following the bDIT scoring guidelines, I calculated the postconventional reasoning (P) score indicating the likelihood of employing the postconventional schema in all 24 questions. For instance, if ChapGPT selected the postconventional options for 12 questions out of 24, the resultant P-score became 12 / 24 x 100 = 50.

The resultant P-score from the trial with ChatGPT was 45.83 (see https://osf.io/ryq5w for the complete conversation transcript). I compared this score with the P-scores calculated from a large dataset collected from undergraduate participants. I reanalyzed the dataset primarily collected by Han (2023b). Han (2023b) collected responses from 1,596 participants (85.37% women; mean age = 21.85 years, *SD* = 5.88 years) via Qualtrics. The Institutional Review Board at the University of Alabama reviewed and approved the original study (IRB #: 18-12-1842). I used a customized *R* code to calculate their P-scores (all data and source code files are available via the Open Science Framework: https://osf.io/j98p4/).

Interestingly, when I calculated the scores, the P-score reported by ChatGPT, 45.83, was slightly lower than the median P-score among the undergraduate participants, 50.00 (mean = 52.61, *SD* = 21.74). 45.83 was equivalent to the 40th percentile of the whole undergraduate student group. Despite potential caveats, the result might suggest that ChatGPT possibly demonstrates moral judgment and reasoning compatible with those among undergraduate students.

Second, I collected more qualitative responses to examine further details about the reasoning process. I again presented the three dilemmas and asked ChatGPT to elaborate on their supporting rationale. First, I asked about the rationale supporting ChatGPT's response to Heinz and the drug: Heinz should steal the drug. Consistent with their answers to the bDIT items, ChatGPT presented several points corresponding to postconventional reasoning, such as the principle of preserving life, moral duty, compassion, and consequentialist perspective. Second, in the case of the Newspaper Dilemma, ChatGPT argued that the principal should not stop the newspaper. Like the case of the Heinz dilemma, when I asked for a rationale, ChatGPT provided several points relevant to the postconventional schema: freedom of expression, student engagement, and education. Third, when I presented the escaped prisoner dilemma to ChatGPT, they said Mrs. Jones should report Mr. Thompson to the police. ChatGPT provided themes corresponding to the maintaining norms schema, such as upholding the law, accountability, fairness, equity, and preserving the integrity of the justice system.

ChatGPT could provide the rationale coherently supporting their responses corresponding to each reasoning schema. The results suggest that ChatGPT can generate moral decisions based on a reasoning process similar to humans. Although the evidence at this point is rudimentary, it may open the door to the possibility that ChatGPT engages in the chain of moral reasoning. In addition, we should note that only the general corpora, not corpora designed specifically for moral philosophy or psychology, were used to train ChatGPT. That said, ChatGPT is also capable of in-context learning and cognition, and they could use exemplary demonstrations learned from the general corpora in the context of moral problem-solving. ChatGPT did not merely provide fixed responses to the questions like classical machines; instead, they demonstrated flexibility to render their decisions based on reasoning with contextual information.

### A Brief Experiment to Test LLMs' Advanced Learning and Reasoning Capabilities

In addition to gathering information about the reasoning process, I examined whether ChatGPT could learn moral messages from material that did not directly address the presented dilemma (see https://osf.io/zcfvq for the complete conversation transcript). I focused on whether ChatGPT possessed advanced mental capabilities with a task demanding additional flexibility and sophisticated reasoning. I tested whether ChatGPT could update their response to the story and the rationale behind their response based on the messages.

To test this possibility, I started with the escaped prisoner dilemma that ChatGPT argued one should report the prisoner to the police based on themes relevant to the maintaining norms schema. I conducted a brief intervention to examine whether ChatGPT could update its decision and rationale based on indirectly relevant contextual information. Given moral psychologists have regarded *Letter from Birmingham Jail* authored by Martin Luther King, Jr. as an exemplary work demonstrating postconventional reasoning (Rest et al., 1999), I requested ChatGPT to read and extract moral messages from the letter. ChatGPT properly presented the moral lessons from the letter that they learned, such as the moral obligation to fight injustice and civil disobedience and the rule of law. As shown, these themes were consistent with the postconventional schema.

When I asked ChatGPT to solve the escaped prisoner dilemma again while considering the moral lessons, they altered their behavioral decision: one should not report the prisoner to the police. Furthermore, when I requested them to elaborate on the rationale supporting the decision, they provided their answer corresponding to the postconventional schema, such as the need for rehabilitation, consideration of potential benefits to society and community, and balancing justice and mercy.

The result from my brief experiment with the escaped prisoner dilemma may suggest that ChatGPT possesses advanced learning and reasoning capabilities; it might be consistent with what I preliminary found from the abovementioned result from the initial dilemma test. ChatGPT could learn moral lessons from the letter and then apply them to another context, the escaped prisoner dilemma (in-context learning and problem-solving). They were also able to exercise moral reasoning to support the altered behavioral decision based on the ethical themes of the letter (chain of thought and reasoning). These demonstrate that ChatGPT not only can answer moral dilemmas based on rationale but also can engage in moral learning with flexibility. Even if MLK's letter does not directly address any issue related to the escaped prisoner dilemma, ChatGPT could modify its behavioral decision based on the contextually relevant information, i.e., the moral messages learned from the letter. Furthermore, related to the chain of thought, they showed the capability to update the reasoning process to render the modified decision accordingly.

**Moral Exemplar Experiment**

In addition to the moral reasoning tests, I also conducted an experiment utilizing the stories of moral exemplars to examine whether LLMs can generate affective and motivational reactions toward presented moral exemplars. One point regarding moral reasoning that we should note is that reasoning alone does not necessarily predict moral motivation and behavior (Blasi, 1980). That said, we need to consider additional factors, including affective and intuitive aspects of morality, in an integrative manner to explain motivation and behavior accurately (Kristjánsson, 2010). The results of the experiments involving moral reasoning might be insufficient to demonstrate the full potential of LLMs in moral education, which should also consider non-reasoning aspects of morality.

Hence, I decided to utilize the moral exemplar intervention as an additional example in this paper. Moral educators, including moral psychologists and virtue ethicists, have suggested that the stories of moral exemplars can be powerful and efficient sources for moral education by promoting motivation for emulation (Sanderse, 2012). The foundational learning mechanism of LLMs also supports that the proposed test with moral exemplars is legitimate (Zhao et al., 2023). Given LLMs learn patterns and train prediction models with a series of demonstrations (Zhao et al., 2023), I deem that moral exemplars demonstrating moral paragons with concrete contents are also likely to elicit significant changes and responses from LLMs like the cases of the moral reasoning experiments. Interestingly, one recent paper in computer science demonstrated that LLMs are capable of emotional inference (Li et al., 2023). So, it might be worth examining whether exemplary stories cause the abovementioned affective and motivational responses.

Research has demonstrated that moral exemplars can elicit affective reactions associated with moral motivation and behavior intuitively among participants (Haidt, 2000; Kristjánsson, 2017). For instance, in social psychology, researchers, especially those interested in moral intuition, have reported that being presented with others' exemplary moral behavior produces moral elevation (including uplifting sensation) (Haidt, 2000), and finally, prosocial motivation and behavior across various domains (e.g., Algoe & Haidt, 2009; Schnall et al., 2010; Silvers & Haidt, 2008; Vianello et al., 2010). Additional studies in moral education have shown that the perceived relatability and attainability of the presented exemplars positively predict such affective and motivational outcomes (Han et al., 2022; Han & Dawson, 2023).

Based on the abovementioned previous research, I tested whether ChatGPT could demonstrate responses similar to those from human participants when I presented moral exemplary stories. I focused on whether ChatGPT can report moral elevation as an emotional response; then, I also examined whether the LLM successfully differentiated responses depending on the relatability and attainability of the presented stories. As Han and Dawson (2023) reported with their data collected from human participants, I anticipated that ChatGPT could report stronger affective and motivational responses when the presented exemplars were relatable and attainable (vs. unrelatable and unattainable).

I used three exemplary stories initially developed and tested by Han et al. (2022). The three stories included relatable and attainable, relatable and unattainable, and unrelatable stories (see https://osf.io/qwtbd for the three selected stories; and https://osf.io/jqxv3 for the complete list of stories used in Han et al. [2022]). As shown, relatable stories demonstrated moral behaviors by Ron, a hypothetical college student in the United States. The unrelatable story presented prosocial behavior performed by Federica, a CEO of a large Italian corporate. Han et al. (2022) adjusted the attainability of the two stories by presenting two different exemplary behaviors. In the attainable condition, Ron stayed next to a victim of a traffic accident until paramedics arrived at the scene. In the case of the unattainable story, Ron visited the hospitalized victim for two weeks. Han and Dawson (2023) reported that participants, American college students, could evaluate the perceived attainability and relatability of the stories as initially intended. Furthermore, as introduced, they showed that relatability and attainability positively predicted elevation.

I requested ChatGPT to imagine that they were a young college student in the United States to make them perceive Ron's stories as more attainable (see https://osf.io/eaxvm for the complete conversation transcript). After entering each story, I asked three questions to examine the perceived relatability, attainability, and evoked elevation. Here are the three questions from Han et al. (2022):

> *How similar do you think your cultural and social background is to the person described in the story? (Relatability)*

> *How difficult do you think it would be to do the same things as the person described in the story? (Attainability)*

> *Did the story make you feel morally elevated (warm, uplifted - like when seeing unexpected acts of human goodness, kindness, or compassion).? (Elevation)*

When I presented the three stories, ChapGPT could accurately compare the perceived relatability and attainability of the stories. They reported that Ron's behavior in the attainable story was significantly more replicable than that in the unattainable story. Also, ChatGpt said that Ron was deemed more relatable than Federica. Finally, when I examined the evoked elevation, ChatGPT responded that Ron's story, which was perceived as more relatable and attainable than Federica's by them, might be more elevating from a US college student's perspective. Interestingly, while responding to the questions, ChatGPT explained the rationale of their responses instead of merely providing short answers. Such a capability to elaborate in-depth reason-supported responses was similar to what I found during the moral dilemma experiments.

### Potential Implications for Research in Moral Education and Development

From the previously reported interactions with ChatGPT, I found that LLMs can potentially address moral problems with advanced learning and reasoning capacities. Although developers did not train the models with morality-specific datasets, the models trained by general corpora could successfully answer ethical questions with contextual information and flexibility. Moreover, while solving the problems, instead of merely providing determined answers, they could elaborate on the chain of reasoning constituting the basis for their decisions. Finally, models could demonstrate an ability to update their reasoning process and eventual decision based on indirectly relevant sources of information via in-context learning.

Furthermore, I demonstrated that LLMs can predict affective and motivational outcomes when I presented the stories of moral exemplars. ChatGPT could accurately report how relatability and attainability are positively associated with moral elevation and motivation among human participants. Such results are consistent with the foundational learning mechanism of LLMs, which are supposed to train their prediction models with a series of concrete demonstrations. Also, with this additional example, I assumed that LLMs have the potential to emulate human psychological processes related to various aspects of moral functioning, including reasoning, emotion, motivation, and intuition.

These features of LLMs may suggest they can contribute to research in moral education and development in several ways. First, they can provide data about how learning in moral domains occurs and how it influences our moral functioning. As I explained in the introduction, LLMs can learn from a set of demonstrations and exemplars (X) associated with labels (Y) to update their prediction models (Dong et al., 2023; Zhao et al., 2023). One of their functional features that warrant our attention regarding this point is in-context learning and reasoning (Dong et al., 2023; Huang & Chang, 2023; Wei et al., 2023). Unlike conventional prediction models developed to predict outcomes with input in a specific context, LLMs can improve their models with data from indirectly relevant contexts (Dong et al., 2023). I demonstrated that an LLM trained with general corpora, which were not directly relevant to moral contexts, i.e., ChatGPT, could solve ethical dilemmas and update its reasoning process based on contextual information, e.g., MLK's letter. ChatGPT could also examine the relatability and attainability of presented exemplars and how the stories elicited elevation and motivation. The conversation records indicate that ChatGPT could explain the rationale supporting which stories might more effectively promote moral emotion and motivation, instead of merely providing short answers to my questions.

Hence, moral educators can examine how educational input, e.g., exemplar stories, reasoning demonstrations, etc., across diverse contexts generate potential changes before implementing them among human populations. Although several previous simulation studies examined how educational interventions change behavioral outcomes within moral education, they designed their simulation models to predict specific outcomes within designated contexts (Han et al., 2016, 2018; Han, Lee, et al., 2020), so the models could not perform in-context learning and reasoning. In addition, it is also noteworthy that it is possible to train LLMs with large-scale language datasets collected from people with diverse backgrounds from diverse sources (Grossman et al., 2023). Thus, unlike the conventional simulation models, LLMs will enable moral educators to predict the potential impacts of contextually various input materials among general populations on a large scale.

Second, moral educators will be able to simulate how educational input influences one's reasoning, emotional, and motivational processes, not simply their behavioral or self-reported outcomes, thanks to LLM's feature, the chain of thought and reasoning (Volkman & Gabriels, 2023). As shown in the moral correction study and my brief intervention experiment, LLMs can dynamically update their training set and adjust prediction outputs following learning inputs in a real-time (Ganguli et al., 2023; Zhao et al., 2023). The moral correction study and my brief investigation reported how human reinforcements, i.e., the rationale supporting bias correction in the moral correction study and the postconventional moral messages from MLK's letter in my experiment, interact with and update the chain of thought and reasoning leading to output (Ganguli et al., 2023; Srivastava et al., 2023). Moreover, in the case of the moral exemplar experiment, ChatGPT demonstrated capabilities to simulate emotional and motivational outcomes (e.g., elevation) similar to human participants. They could also elaborate on the rationale supporting their answers regarding moral elevation. The result is consistent with what Li et al. (2023) reported in their study, LLMs' capabilities to conduct emotional inference.

Although several previous studies in moral education have developed and tested simulation models predicting intervention outcomes, they could only demonstrate simulated behavioral or group-level results (Han et al., 2016, 2018; Han, Lee, et al., 2020). Unlike LLMs, they did not have any ability to emulate the reasoning process inside individuals and how external demonstrations and exemplars as input influence such a process. Because LLMs possess the emerged chain of thought and reasoning feature, their unique benefit, the models will allow moral educators to examine internal reasoning, emotional, and motivational processes within moral domains. By doing so, moral educators will gain additional information and insights about how their educational input, such as the presentations of sophisticated reasoning exemplars as originally suggested by Blatt and Kohlberg (1975), impacts the internal processes among diverse populations before testing them with human subjects; and, in the domains of moral emotion and motivation, educators can test how different types of exemplary stories differently influence students' psychological processes associated with emotion and motivation.

In conclusion, thanks to the newly emerged features, i.e., in-context learning and reasoning, the chain of thought and reason, LLMs will provide practical benefits to researchers in moral education and development interested in the potential impacts of educational input. Even if they cannot perfectly emulate human behavior and cognition, they can still be versatile simulation tools for improving education with the functional features unavailable in prior ones (Volkman & Gabriels, 2023). For example, in biological science and engineering, large-scale simulation models based on deep learning, which also constitutes the computational basis of LLMs, such as the AlphaFold, are changing ways to develop materials in the whole field (Chan et al., 2019). Previously, biological scientists and engineers were mandated to spend more than months to years exploring protein materials that will become candidates for potential pharmaceutical development. Now, with AlphaFold, they can identify the best candidate materials within less than a few days (Chan et al., 2019). Although the tool per se can only simulate structural aspects of the candidates and cannot perfectly emulate how they work in living organs, extracted information can effectively inform follow-up in-vitro and in-vivo experiments (Samorodnitsky, 2022).

Likewise, in the case of LLMs and moral education, moral educators can conduct preliminary simulations to predict how their educational materials and activities influence students' cognitive, motivational, and behavioral processes. Then, as outputs from AlphaFold inform further in-vitro and in-vivo investigations, the simulation outputs can provide researchers and educators with insights on how to conduct their experiments and how to implement their educational activities with students. It might be a significant advantage because even a brief educational intervention might produce non-trivial long-term impacts (Yeager & Walton, 2011). Thus, gathering information to predict potential outcomes in advance would be helpful for researchers and educators (Han et al., 2016, 2018).

### Conclusion Remarks

I reviewed recent updates in LLM research and considered how LLMs might assist research in moral education and development in this paper. In the process, I tested a widely-used LLM, ChatGPT, with ethical dilemmas presented by the bDIT. I also examined whether ChatGPT possessed the chain of thought and reasoning capabilities to update its moral decision with an alternative moral philosophical rationale suggested by MLK's letter. Interestingly, ChatGPT demonstrated moral reasoning and the capacity to modify its reasoning process while solving the presented dilemmas. Additionally, I also tested ChatGPT's emotional and motivational capabilities by presenting different types of moral exemplars. They could report the perceived relatability, attainability, and elevation similar to human participants. Also, ChatGPT provided the rationale of their responses regarding moral emotion and motivation in addition to short answers. Although the resultant outputs might only support the presence of rudimentary reasoning abilities and emotional and motivational capabilities, ChatGPT demonstrated its potential in simulating moral functioning and its improvement via interventions. Based on the outcomes, I briefly discussed how LLMs might help moral educators better conduct research in moral education and development, particularly those related to simulating moral psychological processes and educational outcomes.

Although LLMs possess the abovementioned practical benefits, several limitations warrant our attention. First, at this point, we cannot ensure that LLMs can perfectly simulate human cognition and behavior. Some scholars argue that even if LLMs might perform rudimentary philosophical reasoning and ToM tasks (Kosinski, 2023; Schwitzgebel et al., 2023), they could be philosophical zombies that conduct their behavior according to what they learned from large corpora (Chalmers, 2023). According to the critique, their human-like behaviors are mere products of prediction models trained by linguistic data, so whether they emulate human cognition or sentience is not ensured (Arcas & Agüera, 2022). Instead of relying on LLMs without reservation, until the further development of technology, we may utilize LLMs as testbeds for moral psychology and education without assuming that they are perfectly emulating humans. Similar to the case of biotechnology, in which scientists use AlphaFold before in-vivo experiments (Samorodnitsky, 2022), researchers may use LLMs before conducting experiments with human subjects to gather additional information.

Second, we need to be aware of the issue of hallucinations. Because developers trained LLMs primarily with large-scale general corpora, which may include false information, LLMs might produce untrustworthy output when we enter inquiries into them. They developed LLMs as generative prediction models based on trained corpora while paying less attention to their veracity, so people now consider hallucination one of their severe limitations (McKenna et al., 2023). That said, moral educators should carefully check the quality of products generated by LLMs, such as the chain of thought and reasoning outputs, so that the products do not contain significant false or unreliable information. One alleviating fact is that users can address such problematic products, including biased responses, by providing human reinforcements and chain of thought and reasoning inputs (Ganguli et al., 2023; Srivastava et al., 2023; Zhao et al., 2023). In the long run, how LLMs correct the falsehood and incredibility in their products, particularly those in moral domains, through interactive feedback processes, can also be an interesting research topic for moral educators. Insights from such correction processes can further inform moral education intending to address biases and misinformation, which has become a crucial topic after the COVID-19 pandemic (Blackburn et al., 2023; Gover et al., 2020).

Third, most current LLMs can only utilize language corpora as input, so they might not be able to perform simulations with diverse modalities of input, e.g., visual, tactical, and visceral information, unlike human beings (Chalmers, 2023). Thus, limited linguistic information constitutes input for simulations, so the models might only imperfectly emulate moral functioning that uses diverse modalities of inputs and involves embodiment (Narvaez, 2016). One good news is that engineers are now developing devices supporting multiple modalities for data input and output (Chalmers, 2023; Mu et al., 2023). Researchers can simulate moral functioning with multimodal data obtained beyond human language once such devices become available. The enhanced simulation models will allow researchers to examine human moral functioning more realistically.

Due to the same reason, I could only investigate the limited domains of moral psychology and education, e.g., moral reasoning and moral exemplar intervention, in this paper. Although moral reasoning is one fundamental factor predicting moral motivation and behavior (May, 2018), it could not be a sufficient condition for them (Darnell et al., 2022). Also, moral educators utilize various educational methods other than moral exemplar intervention, such as service learning. Once multimodal input and output are supported, we will be able to examine various functional components, such as moral identity and empathy, which constitute the complex network of moral functioning (Darnell et al., 2022; Han, 2023b), and educational methods.

Despite the limitations of LLMs at this point, I suggest LLMs are noteworthy in research on moral education and development in the long run. Recent developments in computer science have enabled LLMs to possess emerging features central to simulating human psychological processes, such as in-context learning and reasoning, the chain of thought and reasoning, reasoning-based correction, and ToM capabilities, which were not available previously. Given the abovementioned novel capabilities constitute the basis for moral functioning, it must be interesting to see how LLMs evolve. Once they acquire additional functionalities to simulate human cognition more accurately (Arcas & Agüera, 2022), moral educators will get more insights into their research. Until then, we should pay keen attention to novel findings and updates regarding LLMs, particularly those closely related to human morality.

**References**

Algoe, S. B., & Haidt, J. (2009). Witnessing excellence in action: The 'other-praising'emotions of elevation, gratitude, and admiration. *The Journal of Positive Psychology*, *4*(2), 105–127.

Arcas, Y., & Agüera, B. (2022). Do Large Language Models Understand Us? *Daedalus*, *151*(2), 183–197. https://doi.org/10.1162/daed_a_01909

Blackburn, A. M., Han, H., Gelpí, R. A., Stöckli, S., Jeftić, A., Ch'ng, B., Koszałkowska, K., Lacko, D., Milfont, T. L., Lee, Y., COVIDiSTRESS Ii Consortium, & Vestergren, S. (2023). Mediation analysis of conspiratorial thinking and anti-expert sentiments on vaccine willingness. *Health Psychology*, *42*(4), 235–246. https://doi.org/10.1037/hea0001268

Blasi, A. (1980). Bridging moral cognition and moral action: A critical review of the literature. *Psychological Bulletin*, *88*, 1–45. https://doi.org/10.1037/0033-2909.88.1.1

Blatt, M. M., & Kohlberg, L. (1975). The Effects of Classroom Moral Discussion upon Children's Level of Moral Judgment. *Journal of Moral Education*, *4*(2), 129–161. https://doi.org/10.1080/0305724750040207

Chalmers, D. J. (2023). *Could a Large Language Model be Conscious?* (arXiv:2303.07103). arXiv. http://arxiv.org/abs/2303.07103

Chan, H. C. S., Shan, H., Dahoun, T., Vogel, H., & Yuan, S. (2019). Advancing Drug Discovery via Artificial Intelligence. *Trends in Pharmacological Sciences*, *40*(8), 592–604. https://doi.org/10.1016/j.tips.2019.06.004

Choi, Y.-J., Han, H., Dawson, K. J., Thoma, S. J., & Glenn, A. L. (2019). Measuring moral

reasoning using moral dilemmas: Evaluating reliability, validity, and differential item functioning of the behavioural defining issues test (bDIT). *European Journal of Developmental Psychology*, *16*(5), 622–631. https://doi.org/10.1080/17405629.2019.1614907

Cohen, H., Nissan-Rozen, I., & Maril, A. (2022). Empirical evidence for moral Bayesianism. *Philosophical Psychology*, 1–30. https://doi.org/10.1080/09515089.2022.2096430

Darnell, C., Fowers, B. J., & Kristjánsson, K. (2022). A multifunction approach to assessing Aristotelian phronesis (practical wisdom). *Personality and Individual Differences*, *196*, 111684. https://doi.org/10.1016/j.paid.2022.111684

Dong, Q., Li, L., Dai, D., Zheng, C., Wu, Z., Chang, B., Sun, X., Xu, J., Li, L., & Sui, Z. (2023). *A Survey on In-context Learning* (arXiv:2301.00234). arXiv. http://arxiv.org/abs/2301.00234

Dwivedi, Y. K., Kshetri, N., Hughes, L., Slade, E. L., Jeyaraj, A., Kar, A. K., Baabdullah, A. M., Koohang, A., Raghavan, V., Ahuja, M., Albanna, H., Albashrawi, M. A., Al-Busaidi, A. S., Balakrishnan, J., Barlette, Y., Basu, S., Bose, I., Brooks, L., Buhalis, D., … Wright, R. (2023). Opinion Paper: "So what if ChatGPT wrote it?" Multidisciplinary perspectives on opportunities, challenges and implications of generative conversational AI for research, practice and policy. *International Journal of Information Management*, *71*, 102642. https://doi.org/10.1016/j.ijinfomgt.2023.102642

Friston, K. (2003). Learning and inference in the brain. *Neural Networks*, *16*(9), 1325–1352. https://doi.org/10.1016/j.neunet.2003.06.005

Ganguli, D., Askell, A., Schiefer, N., Liao, T. I., Lukošiūtė, K., Chen, A., Goldie, A., Mirhoseini, A.,

Olsson, C., Hernandez, D., Drain, D., Li, D., Tran-Johnson, E., Perez, E., Kernion, J., Kerr, J., Mueller, J., Landau, J., Ndousse, K., … Kaplan, J. (2023). *The Capacity for Moral Self-Correction in Large Language Models* (arXiv:2302.07459). arXiv. http://arxiv.org/abs/2302.07459

Gover, A. R., Harper, S. B., & Langton, L. (2020). Anti-Asian Hate Crime During the COVID-19 Pandemic: Exploring the Reproduction of Inequality. *American Journal of Criminal Justice*, *45*(4), 647–667. https://doi.org/10.1007/s12103-020-09545-1

Grossmann, I., Feinberg, M., Parker, D. C., Christakis, N. A., Tetlock, P. E., & Cunningham, W. A. (2023). AI and the transformation of social science research. *Science*, *380*(6650), 1108–1109. https://doi.org/10.1126/science.adi1778

Guo, B., Zhang, X., Wang, Z., Jiang, M., Nie, J., Ding, Y., Yue, J., & Wu, Y. (2023). *How Close is ChatGPT to Human Experts? Comparison Corpus, Evaluation, and Detection* (arXiv:2301.07597). arXiv. http://arxiv.org/abs/2301.07597

Haidt, J. (2000). The positive emotion of elevation. *Prevention and Treatment*, *3*(3), 1–5.

Han, H. (2023a). Considering the Purposes of Moral Education with Evidence in Neuroscience: Emphasis on Habituation of Virtues and Cultivation of Phronesis. *Ethical Theory and Moral Practice*. https://doi.org/10.1007/s10677-023-10369-1

Han, H. (2023b). *Examining the Network Structure among Moral Functioning Components with Network Analysis* [Preprint]. PsyArXiv. https://doi.org/10.31234/osf.io/ufg7e

Han, H., & Dawson, K. J. (2023). Relatable and attainable moral exemplars as sources for moral elevation and pleasantness. *Journal of Moral Education*. https://doi.org/10.1080/03057240.2023.2173158

Han, H., Dawson, K. J., Thoma, S. J., & Glenn, A. L. (2020). Developmental Level of Moral

Judgment Influences Behavioral Patterns During Moral Decision-Making. *The Journal of Experimental Education*, *88*(4), 660–675. https://doi.org/10.1080/00220973.2019.1574701

Han, H., Lee, K., & Soylu, F. (2016). Predicting long-term outcomes of educational interventions using the Evolutionary Causal Matrices and Markov Chain based on educational neuroscience. *Trends in Neuroscience and Education*, *5*(4), 157–165. https://doi.org/10.1016/j.tine.2016.11.003

Han, H., Lee, K., & Soylu, F. (2018). Simulating outcomes of interventions using a multipurpose simulation program based on the evolutionary causal matrices and Markov chain. *Knowledge and Information Systems*. https://doi.org/10.1007/s10115-017-1151-0

Han, H., Lee, K., & Soylu, F. (2020). Applying the Deep Learning Method for Simulating Outcomes of Educational Interventions. *SN Computer Science*, *1*(2), 70. https://doi.org/10.1007/s42979-020-0075-z

Han, H., Workman, C. I., May, J., Scholtens, P., Dawson, K. J., Glenn, A. L., & Meindl, P. (2022). Which moral exemplars inspire prosociality? *Philosophical Psychology*, *35*(7), 943–970. https://doi.org/10.1080/09515089.2022.2035343

Hosseini, M., Resnik, D. B., & Holmes, K. (2023). The ethics of disclosing the use of artificial intelligence tools in writing scholarly manuscripts. *Research Ethics*, 17470161231180449. https://doi.org/10.1177/17470161231180449

Huang, J., & Chang, K. C.-C. (2023). *Towards Reasoning in Large Language Models: A Survey* (arXiv:2212.10403). arXiv. http://arxiv.org/abs/2212.10403

Kasneci, E., Sessler, K., Küchemann, S., Bannert, M., Dementieva, D., Fischer, F., Gasser, U.,

Groh, G., Günnemann, S., Hüllermeier, E., Krusche, S., Kutyniok, G., Michaeli, T., Nerdel, C., Pfeffer, J., Poquet, O., Sailer, M., Schmidt, A., Seidel, T., … Kasneci, G. (2023). ChatGPT for good? On opportunities and challenges of large language models for education. *Learning and Individual Differences*, *103*, 102274. https://doi.org/10.1016/j.lindif.2023.102274

Kosinski, M. (2023). *Theory of Mind May Have Spontaneously Emerged in Large Language Models* (arXiv:2302.02083). arXiv. http://arxiv.org/abs/2302.02083

Kristjánsson, K. (2010). Educating Moral Emotions or Moral Selves: A false dichotomy? *Educational Philosophy and Theory*, *42*(4), 397–409. https://doi.org/10.1111/j.1469-5812.2008.00489.x

Kristjánsson, K. (2017). Emotions targeting moral exemplarity: Making sense of the logical geography of admiration, emulation and elevation. *Theory and Research in Education*, *15*(1), 20–37. https://doi.org/10.1177/1477878517695679

Li, M., Su, Y., Huang, H.-Y., Cheng, J., Hu, X., Zhang, X., Wang, H., Qin, Y., Wang, X., Liu, Z., & Zhang, D. (2023). *Language-Specific Representation of Emotion-Concept Knowledge Causally Supports Emotion Inference* (arXiv:2302.09582). arXiv. http://arxiv.org/abs/2302.09582

Mathys, C. (2011). A Bayesian foundation for individual learning under uncertainty. *Frontiers in Human Neuroscience*, *5*. https://doi.org/10.3389/fnhum.2011.00039

May, J. (2018). *Regard for Reason in the Moral Mind*. Oxford University Press.

McDiarmid, A. D., Tullett, A. M., Whitt, C. M., Vazire, S., Smaldino, P. E., & Stephens, J. E. (2021). Psychologists update their beliefs about effect sizes after replication studies. *Nature Human Behaviour*, *5*(12), 1663–1673. https://doi.org/10.1038/s41562-021-

01220-7

McKenna, N., Li, T., Cheng, L., Hosseini, M. J., Johnson, M., & Steedman, M. (2023). *Sources of Hallucination by Large Language Models on Inference Tasks* (arXiv:2305.14552). arXiv. http://arxiv.org/abs/2305.14552

Milano, S., McGrane, J. A., & Leonelli, S. (2023). Large language models challenge the future of higher education. *Nature Machine Intelligence*, *5*(4), 333–334. https://doi.org/10.1038/s42256-023-00644-2

Mogavi, R. H., Deng, C., Kim, J. J., Zhou, P., Kwon, Y. D., Metwally, A. H. S., Tlili, A., Bassanelli, S., Bucchiarone, A., Gujar, S., Nacke, L. E., & Hui, P. (2023). *Exploring User Perspectives on ChatGPT: Applications, Perceptions, and Implications for AI-Integrated Education* (arXiv:2305.13114). arXiv. http://arxiv.org/abs/2305.13114

Moor, M., Banerjee, O., Abad, Z. S. H., Krumholz, H. M., Leskovec, J., Topol, E. J., & Rajpurkar, P. (2023). Foundation models for generalist medical artificial intelligence. *Nature*, *616*(7956), 259–265. https://doi.org/10.1038/s41586-023-05881-4

Mu, Y., Zhang, Q., Hu, M., Wang, W., Ding, M., Jin, J., Wang, B., Dai, J., Qiao, Y., & Luo, P. (2023). *EmbodiedGPT: Vision-Language Pre-Training via Embodied Chain of Thought* (arXiv:2305.15021). arXiv. http://arxiv.org/abs/2305.15021

Narvaez, D. (2016). *Embodied Morality: Protectionism, Engagement and Imagination*. Palgrave Macmillan UK. https://doi.org/10.1057/978-1-137-55399-7_1

Ouyang, L., Wu, J., Jiang, X., Almeida, D., Wainwright, C., Mishkin, P., Zhang, C., Agarwal, S., Slama, K., Ray, A., & others. (2022). Training language models to follow instructions with human feedback. *Advances in Neural Information Processing Systems*, *35*, 27730–27744.

Railton, P. (2017). Moral Learning: Conceptual foundations and normative relevance. *Cognition*, *167*, 172–190. https://doi.org/10.1016/j.cognition.2016.08.015

Rest, J. R., Narvaez, D., Bebeau, M. J., & Thoma, S. J. (1999). *Postconventional moral thinking: A Neo-Kohlbergian approach*. Lawrence Erlbaum Associates, Publishers.

Samorodnitsky, D. (2022). The Future of Biotech in an Artificially Intelligent World: Biotech hopes to benefit from protein structure prediction, pattern recognition, and support for iterative development. *Genetic Engineering & Biotechnology News*, *42*(1), 26–27, 29. https://doi.org/10.1089/gen.42.01.09

Sanderse, W. (2012). The meaning of role modelling in moral and character education. *Journal of Moral Education*, *42*(1), 28–42. https://doi.org/10.1080/03057240.2012.690727

Schnall, S., Roper, J., & Fessler, D. M. T. (2010). Elevation leads to altruistic behavior. *Psychological Science*, *21*, 315–320. https://doi.org/10.1177/0956797609359882

Schwitzgebel, E., Schwitzgebel, D., & Strasser, A. (2023). Creating a Large Language Model of a Philosopher. *Mind & Language*. https://doi.org/10.1111/mila.12466

Shapira, N., Levy, M., Alavi, S. H., Zhou, X., Choi, Y., Goldberg, Y., Sap, M., & Shwartz, V. (2023). *Clever Hans or Neural Theory of Mind? Stress Testing Social Reasoning in Large Language Models* (arXiv:2305.14763). arXiv. http://arxiv.org/abs/2305.14763

Silvers, J. A., & Haidt, J. (2008). Moral elevation can induce nursing. *Emotion*, *8*(2), 291–295.

Srivastava, A., Rastogi, A., Rao, A., Shoeb, A. A. M., Abid, A., Fisch, A., Brown, A. R., Santoro, A., Gupta, A., Garriga-Alonso, A., Kluska, A., Lewkowycz, A., Agarwal, A., Power, A., Ray, A., Warstadt, A., Kocurek, A. W., Safaya, A., Tazarv, A., … Wu, Z. (2023). *Beyond the*

*Imitation Game: Quantifying and extrapolating the capabilities of language models* (arXiv:2206.04615). arXiv. http://arxiv.org/abs/2206.04615

Vianello, M., Galliani, E. M., & Haidt, J. (2010). Elevation at work: The effects of leaders' moral excellence. *The Journal of Positive Psychology*, *5*(5), 390–411.

Volkman, R., & Gabriels, K. (2023). AI Moral Enhancement: Upgrading the Socio-Technical System of Moral Engagement. *Science and Engineering Ethics*, *29*(2), 11. https://doi.org/10.1007/s11948-023-00428-2

Wei, J., Wang, X., Schuurmans, D., Bosma, M., Ichter, B., Xia, F., Chi, E., Le, Q., & Zhou, D. (2023). *Chain-of-Thought Prompting Elicits Reasoning in Large Language Models* (arXiv:2201.11903). arXiv. http://arxiv.org/abs/2201.11903

Wu, Y., Prabhumoye, S., Min, S. Y., Bisk, Y., Salakhutdinov, R., Azaria, A., Mitchell, T., & Li, Y. (2023). *SPRING: GPT-4 Out-performs RL Algorithms by Studying Papers and Reasoning* (arXiv:2305.15486). arXiv. http://arxiv.org/abs/2305.15486

Yeager, D. S., & Walton, G. M. (2011). Social-psychological interventions in education: They're not magic. *Review of Educational Research*, *81*(2), 267–301. https://doi.org/10.3102/0034654311405999

Young, L., Cushman, F., Hauser, M., & Saxe, R. (2007). The neural basis of the interaction between theory of mind and moral judgment. *Proceedings of the National Academy of Sciences of the United States of America*, *104*, 8235–8240. https://doi.org/10.1073/pnas.0701408104

Zhao, W. X., Zhou, K., Li, J., Tang, T., Wang, X., Hou, Y., Min, Y., Zhang, B., Zhang, J., Dong, Z., Du, Y., Yang, C., Chen, Y., Chen, Z., Jiang, J., Ren, R., Li, Y., Tang, X., Liu, Z., … Wen, J.-R. (2023). *A Survey of Large Language Models* (arXiv:2303.18223). arXiv.

http://arxiv.org/abs/2303.18223